JUNE 2025

# A COMPREHENSIVE GUIDE TO THE U.S. FEDERAL CIVIL SPACE BUDGET

**LINDSAY DEMARCHI**
**THE AEROSPACE CORPORATION**

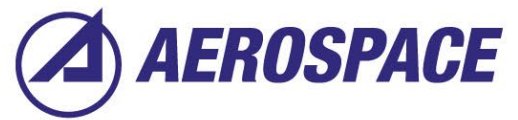

**DR. LINDSAY DEMARCHI**

Dr. Lindsay DeMarchi is a policy analyst in the Center for Space Policy and Strategy at The Aerospace Corporation. Her expertise centers around civil space, interests of the public good, and related intersections and opportunities in a burgeoning commercial space sector. Prior to joining Aerospace, DeMarchi was a congressional fellow in the United States Senate covering space, transportation, and natural resource policy topics. She led the development of legislation on space launch licensing, space debris, light pollution from satellites, and commercial low Earth orbit (LEO) destinations. DeMarchi earned a bachelor's degree in physics and astronomy from Colgate University, a master's degree in physics from Syracuse University, and a master's degree and a doctorate in astronomy from Northwestern University.

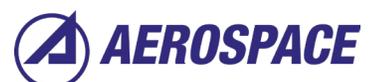

# Summary

This paper presents an overview of the U.S. federal civil space budget between FY23 and proposed spending in FY25,[*] providing a foundational reference dataset for civil space budget analyses. Projects that support, enable, or leverage space activities for civil purposes are found in more than 100 individual line items spread across 17 federal departments and agencies and are funded by 4 different appropriations bills.[†] Across all these elements, civil space-related appropriations total roughly $44 billion in FY25, of which NASA constitutes about 58 percent (Figure 1).[‡]

To better view the cross-cutting nature of space missions and the support that enables them, the second half of this report introduces a novel representation of the civil space budget by organizing the data into six national priority areas:

♦ **American Leadership and Manufacturing**

♦ **Workforce Development**

♦ **Fundamental Science**

♦ **Efficiency, Improvements, and Growth**

♦ **Homeland Security**

♦ **Infrastructure, Energy, and Resiliency**

♦ **Remote Sensing Applications (Figure 2)**

This organization of the data highlights the ubiquity of space-related activity throughout diverse sectors to reach the same common goal.

Given the significant changes underway in the federal government's structure and budget, the Center for Policy and Strategy at The Aerospace Corporation will revise and update this analysis as those changes are finalized, and we welcome feedback from other analysts to refine the categorization. All the budget data discussed in this paper is also available as a separate Excel file at https://csps.aerospace.org/sites/default/files/2025-06/U.S.FederalCivilSpaceBudgetCompanionSpreadsheet_0.zip.

---

[*] Instead of adopting regular appropriations for FY25, Congress enacted a full-year continuing resolution based primarily on FY24 appropriations levels. This paper utilizes FY25 agency requests.

[†] This study defines "civil space" to encompass all activity outside defense or intelligence purposes.

[‡] Please note that, due to the level of detail available in budget requests, some line items used in this estimate are not solely devoted to space-related activity and in other cases line-item values were not found. With the same caveats, the total in FY24 was about $41 billion; the FY23 total was about $43 billion.

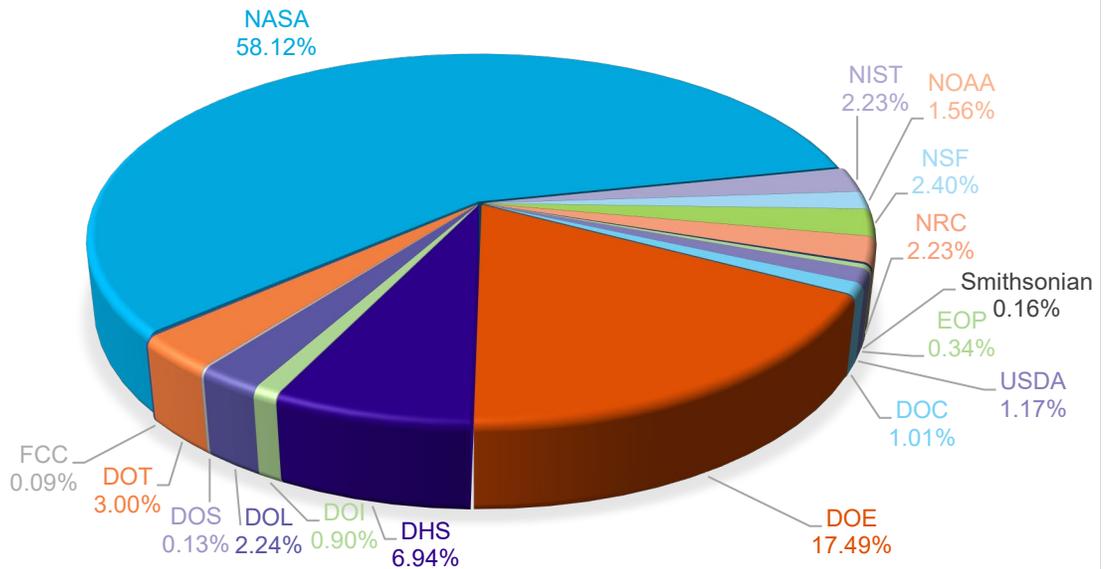

**Figure 1:** *The federal civil space budget for FY25 organized by department and agency. Please note, due to the level of detail available in budget requests, some line items used in this estimate are not solely devoted to space-related activity and, in other cases, line-item values were not found.*

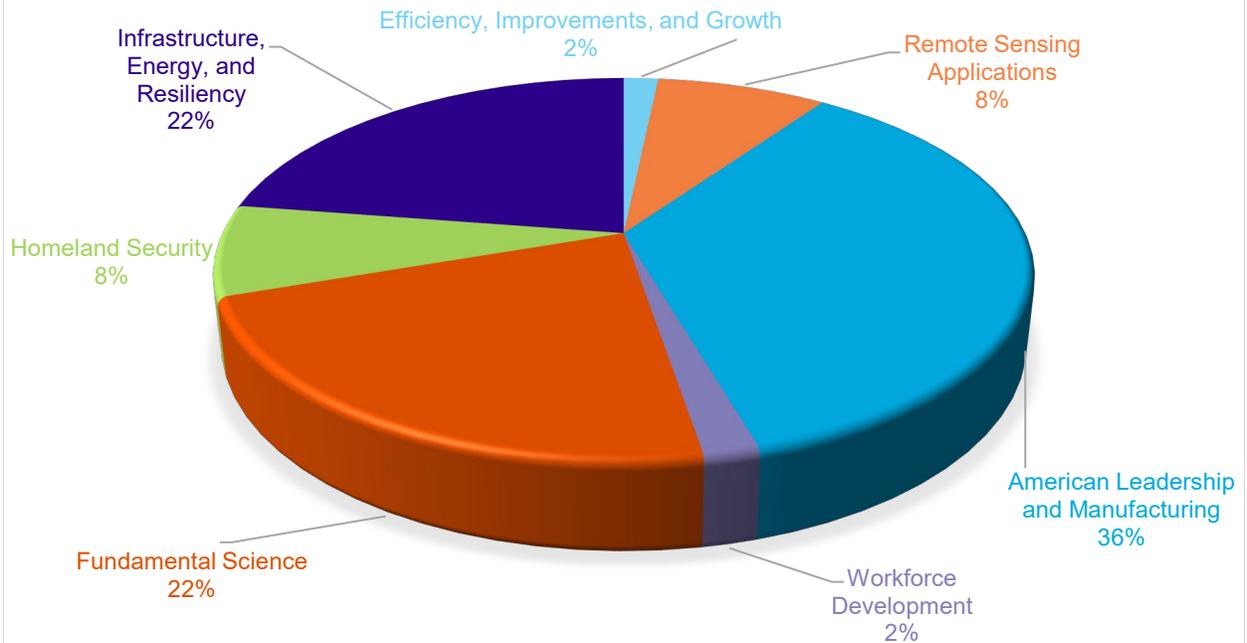

**Figure 2:** *The federal civil space budget for FY25 organized by national priority. Please note, due to the level of detail available in budget requests, some line items used in this estimate are not solely devoted to space-related activity and, in other cases, line-item values were not found.*

# Contents

## Introduction

The term civil space is generally understood among the public and policymakers to refer to non-national security agencies with direct missions to space, such as the National Aeronautics and Space Administration (NASA) and the National Oceanic and Atmospheric Administration (NOAA). Yet, the Department of Transportation (DOT) and the Department of Labor (DOL) are equally inextricable from civil space capabilities because they support the space industry workforce and ensure the safe transportation of critical supply chain materials. The Department of Agriculture (USDA) and the Smithsonian Institution, meanwhile, make extensive use of data that can only be acquired from space in their transdisciplinary public projects. Departments such as the Department of Interior (DOI) and the Department of Energy (DOE), too, rely heavily on space data to achieve their missions, while the Executive Office of the President (EOP), Federal Aviation Administration (FAA), and Federal Communications Commission (FCC) craft policy and regulations that direct and enable space activities.

This paper broadens the definition of civil space to include activities that support, enable, or leverage space through direct and indirect means. The growing number of civil space departments and agencies, as highlighted in Figure 3, is one way of measuring space's increased national importance. The ubiquitous use of space is also a result of capabilities becoming more accessible, allowing department agendas to incorporate space as a tool rather than endeavor a space-centric, long-term, strategic mission.[*] As a result, the emerging civil space enterprise resembles a complex food chain, with embedded interdependencies between agencies.[†] This paper builds upon previous work that has identified some areas of civil space appropriations,[1] defense space appropriations,[2] and individual agency budget trends over time[3, 4] by providing snapshots of each agency's appropriations requests and discussing its line items' relevance to civil space under this new definition.

---

[*] For a comprehensive overview in strategic foresight methodologies applied to space, see https://csps.aerospace.org/papers/project-north-star-strategic-foresight-us-grand-strategy and tools developed in https://csps.aerospace.org/papers/strategic-foresight-space-enterprise
[†] See https://csps.aerospace.org/papers/framing-space-agenda-through-strategic-foresight for applying foresighting to interconnected themes and issues across space priority areas.

# The Government of the United States

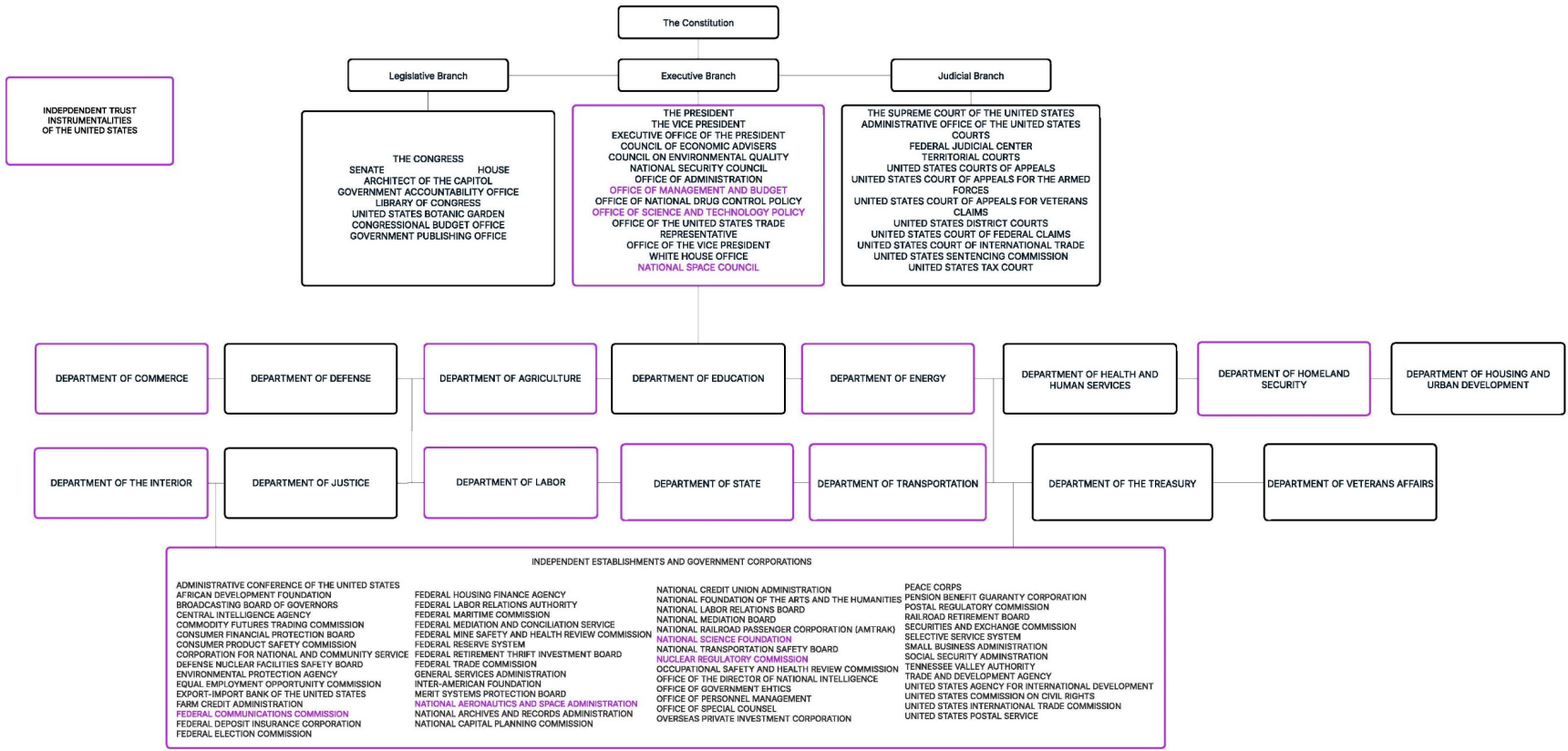

**Figure 3:** *An overview of the organizational structure of the U.S. government. Boxes highlighted in purple indicate departments and independent agencies with programs that are supported, enabled, or leveraged by civil space programs and line items. Chart adapted from the General Services Administration.[5]*

An outcome of space's increased prominence is its relevance in unexpected places that have become vital for maintaining a balance of innovation, quality of life, public safety, and economic prosperity. As a means of identifying this trend, this paper presents a novel view of the federal civil space budget that groups the dataset of more than 100 individual line items into non-space categories associated with current national priorities. These categories are:

- ◆ American Leadership and Manufacturing
- ◆ Workforce Development
- ◆ Fundamental Science
- ◆ Efficiency, Improvements, and Growth
- ◆ Homeland Security
- ◆ Infrastructure, Energy, and Resiliency
- ◆ Remote Sensing

Such a categorization offers a new perspective on the connections between federal investments in space to livelihoods, manufacturing, security, and resources, better reflecting the widespread use of space and its value to the American people.

## *Methodology: Compiling and Interpreting Budgetary Information*

The Office of Management and Budget (OMB) provides some guidance on how agencies should organize their annual President's Budget Requests. However, agencies often include additional information based on mission needs and accumulated guidance from the various appropriations subcommittees reviewing their requests. This results in assorted levels of granularity among all federal budgetary submissions, and direct line-item comparisons between agencies becomes a challenge. The top image in Figure 4 is an example of depth specific to the National Science Foundation (NSF) in its FY24 President's Budget Request, where the NSF denotes individual facilities supporting major projects in addition to the projects' overall proposed funding.

To a similar effect, agencies' requests following continuing resolutions (CRs) have an added layer of diversity in reporting that further complicates direct comparisons of project funding. In response to the CR in 2024, some agencies reported the amount *spent* in 2023 in their FY25 budget request documents, while others recorded the *appropriated* (or "*enacted*") amount for FY23. Still, others reported the "*base*" dollar amounts, or bottom lines necessary to continue base operations (Figure 4).[*] Where possible, this paper footnotes which values are used if they are not reported as "*requested*."

As a result of the variance in reporting among President's Budget Requests, this analysis preferentially reports dollar amounts from appropriations bills and CRs as the common mechanism among diverse sources. Unfortunately, this can be at the expense of detail—the text of congressional spending bills does not go into the same level of detail explicating individual budget line items (Figure 4). As a result, one must infer between the President's Budget Requests and congressional bills how these dollar amounts are distributed to programs in practice within the agencies and departments. Where possible, this analysis references the location of each line item in the respective President's Budget Requests. Some agencies, such as NASA, additionally received emergency funding in 2024.[6] Tracking all related emergency funding is beyond the scope of this paper.

---

[*] In simpler words: Some agencies looked back on the year and reported what they spent, independent of what they had originally asked of Congress. Others reported the amounts they had originally requested from Congress. Additional agencies reported the amounts Congress gave them at the start of the previous fiscal year. Still others reported the bare minimum necessary for their agency to exist.

| | FY 2023 Base Plan | FY 2024 (TBD) | FY 2025 Request | Change over FY 2023 Base Plan | |
|---|---|---|---|---|---|
| **MAJOR FACILITIES FUNDING, BY PROJECT** (Dollars in Millions) | | | | Amount | Percent |
| **Operations and Maintenance of Major Facilities** | **$996.74** | **-** | **$1,120.33** | **$123.59** | **12.4%** |
| National Ecological Observatory Network (NEON) | 71.71 | - | 82.02 | 10.31 | 14.4% |
| ***Biological Sciences*** | ***$71.71*** | **-** | ***$82.02*** | ***$10.31*** | ***14.4%*** |
| Academic Research Fleet[1] | 136.09 | - | 151.33 | 15.24 | 11.2% |
| National Center for Atmospheric Research (NCAR) FFRDC | 116.20 | - | 124.59 | 8.39 | 7.2% |
| National Geophysical Facility (NGF)[2] | 37.92 | - | 45.29 | 7.37 | 19.4% |
| *Geodetic Facility for the Advancement of GEoscience (GAGE)* [2] | *14.55* | - | *8.55* | *-6.00* | *-41.2%* |
| *Seismological Facility for the Advancement of GEoscience (SAGE)* [2] | *23.37* | - | *13.25* | *-10.12* | *-43.3%* |
| *NGF O&M* [2] | - | - | *23.49* | *23.49* | *N/A* |
| Ocean Observatories Initiative (OOI)[1] | 42.02 | - | 47.76 | 5.74 | 13.7% |
| U.S. Sub-seafloor Sampling (S3P) (Formerly IODP) | 50.40 | - | 55.51 | 5.11 | 10.1% |
| ***Geosciences*** | ***$382.63*** | **-** | ***$424.48*** | ***$41.85*** | ***10.9%*** |
| Green Bank Observatory (GBO) FFRDC[3] | 10.83 | - | 9.68 | -1.15 | -10.6% |
| Large Hadron Collider (LHC) - ATLAS and CMS | 20.50 | - | 20.50 | - | - |
| Laser Interferometer Gravitational Wave Observatory (LIGO) | 45.00 | - | 49.00 | 4.00 | 8.9% |
| National High Magnetic Field Laboratory (NHMFL)[3] | 39.91 | - | 39.13 | -0.78 | -2.0% |
| National Radio Astronomy Observatory (NRAO) FFRDC[3] | 93.66 | - | 96.71 | 3.05 | 3.3% |
| *NRAO O&M* [3,4] | *43.03* | - | *43.00* | *-0.03* | *-0.1%* |
| *Atacama Large Millimeter Array (ALMA) O&M* | *50.63* | - | *53.71* | *3.08* | *6.1%* |
| National Solar Observatory (NSO) FFRDC[3] | 26.56 | - | 34.24 | 7.68 | 28.9% |
| *NSO O&M* | *5.88* | - | *6.24* | *0.36* | *6.1%* |
| *Daniel K. Inouye Solar Telescope (DKIST)* [3] | *20.68* | - | *28.00* | *7.32* | *35.4%* |
| NSF's National Optical-Infrared Astronomy Research Laboratory FFRDC[3] | 73.57 | - | 86.40 | 12.83 | 17.4% |
| *NOIRLab O&M (Mid-Scale Observatories & Community Science and Data Center)* [3,5] | *28.49* | - | *24.82* | *-3.67* | *-12.9%* |
| *GEMINI Observatory O&M* | *22.98* | - | *25.49* | *2.51* | *10.9%* |
| *Vera C. Rubin Observatory O&M* | *22.10* | - | *36.09* | *13.99* | *63.3%* |
| ***Mathematical and Physical Sciences*** | ***$310.03*** | **-** | ***$335.66*** | ***$25.63*** | ***8.3%*** |
| Antarctic Facilities and Operations (AFO) | 224.71 | - | 269.94 | 45.23 | 20.1% |
| IceCube Neutrino Observatory (ICNO) | 7.66 | - | 8.23 | 0.57 | 7.4% |
| ***Office of Polar Programs*** | ***$232.37*** | **-** | ***$278.17*** | ***$45.80*** | ***19.7%*** |
| **Major Research Facilities Construction Investments** | **$216.66** | **-** | **$345.00** | **$128.34** | **59.2%** |
| **R&RA Design Stage Activities[6]** | **$30.43** | **-** | **$46.00** | **$15.57** | **51.2%** |
| **Major Research Equipment and Facilities Construction (MREFC)** | **$186.23** | **-** | **$299.00** | **$112.77** | **60.6%** |
| **Total, Major Research Facilities** | **$1,213.40** | **-** | **$1,465.33** | **$251.93** | **20.8%** |

```
         National Science Foundation

Research and related activities 1/.................   6,931,136    8,915,900    7,067,190   +136,054   -1,848,710
    Defense function...............................      90,000      102,000      109,310    +19,310       +7,310
                                                     ------------ ------------ ------------ ---------- ------------
         Subtotal..................................   7,021,136    9,017,900    7,176,500   +155,364   -1,841,400

Major Research Equipment and Facilities Construction..  187,230      304,670          ---   -187,230     -304,670
Major Research Equipment and Facilities Construction
  (emergency)......................................         ---          ---      234,000   +234,000     +234,000
Education and Human Resources 1/...................   1,154,000    1,496,180    1,172,000    +18,000     -324,180
Agency Operations and Award Management.............     448,000      503,870      448,000        ---      -55,870
Office of the National Science Board...............       5,090        5,250        5,090        ---         -160
Office of Inspector General........................      23,393       26,810       24,410     +1,017       -2,400
         Total, National Science Foundation........   8,838,849   11,354,680    9,060,000   +221,151   -2,294,680
                                                     ============ ============ ============ ========== ============
```

***Figure 4:*** *An example of differences in granularity among budget documents from which data has been taken for this study. Top image: an example of budgetary detail in the FY24 NSF President's Budget Request.[7] Bottom image: budgetary detail in the NSF section of the FY24 CR.[8] Note that the level of detail in the President's Budget Request is not present in the final appropriations bill.*

### *How Much of a Line Item Represents Space?*

This reference dataset contains more than 100 individual line items from the total federal budget. The degree to which each budget line item represents space activity varies. For example, the FAA's *Advanced Technology Development and Prototyping* contains some individual space-related projects, while NASA's *Deep Space Exploration Systems* is entirely devoted to space activity. For comprehensiveness and completeness, all line items in this paper are calculated and summed according to the detail offered by appropriations budget requests put forth by each agency. It is beyond the scope of this publication to calculate the budget on a more granular project-by-project basis. However, descriptions of each project's relevancy to civil space are provided in each section.

The intent of this referential dataset is to begin a broader discussion of the direct and indirect relationships civil space budget line items have to broader national priorities. For example, NASA's *Heliophysics* can be seen as fundamental science research or critical to infrastructure resiliency in preparing for solar weather. The reader is encouraged to experiment with other forms of estimating category totals using the spreadsheets published along with this paper.

### *Accounting for Inflation*

A consideration for any budget analysis is whether to account for inflation when presenting historical data. This paper uses *current dollars*, or the actual dollar amount provided that fiscal year, and therefore does not adjust for inflation. To calculate the inflation-corrected value for FY25, please refer to Table 10.1 of the Historical Tables published by the Office of Management and Budget.[9]

### What to Expect in This Paper

The first half of this paper presents the federal civil space budget across departments and agencies that have submitted their own reports to Congress, listed in alphabetical order. Each entry includes the relevant appropriations bill that funds the department or agency, the establishing legislation for the department or agency (not including subsequent changes), a brief description of the department or agency, and its connection to civil space. Some agency section titles are independent agencies, while others are represented separately for submitting their own reports to Congress but technically exist within departments. Those sections denote the department under which the agency lives. Each section contains a table of curated line items that ultimately tally the amounts appropriated from FY23 to the requested amounts of FY25. The totals per table estimate civil space contributions contributed over all line items, except where noted.

The second half of this paper reorganizes the same federal civil space budget information into a set of categories that reflects national priorities and spans multiple departments and agencies or includes different parts of an agency's budget. Each section contains a description of the category, and its tables have an additional column to denote what agency the line item is from. Citations for all values in these sections can be found in their original department or agency tables.

# Civil Space Budget by Department and Agency

## Executive Office of the President (EOP)

*Appropriations bill: Commerce, Justice, and Science (CJS)*
*Establishing legislation: The Reorganization Act of 1939, Reorganization Plan Number 1[10]*

The EOP mission is to provide support to the president. On space, the EOP serves critical leadership roles in civil space, both in orchestrating the budgetary process through the OMB and providing guidance in space policy. The National Space Council and the Office of Science and Technology Policy (OSTP) puts forth presidential policy recommendations, executive orders, and strategies for the executive branch to follow and guide national priorities. In particular, the OSTP crafts policy and analysis for the president related to cutting-edge technological issues such as quantum information, artificial intelligence, biotechnology, and space.

| Line Item | FY25 | FY24 [11] | FY23 [*,12] |
|---|---|---|---|
| **National Space Council** | 2 [13] | 1.965 | 1.965 |
| **Office of Science and Technology Policy (OSTP)** | 8 [14] | 7.965 | 7.965 |
| **OMB** | 138.3 [15] | 128.035 [16] | 128.035 [17] |
| Civil Space Estimate: | **148.3** | **137.695** | **137.695** |

*AMOUNTS IN MILLIONS.*

---

[*] Unless otherwise noted, values are enacted.

## Department of Agriculture (USDA)

*Appropriations bill: Agriculture, Rural Development, Food and Drug Administration, and Related Agencies*
*Establishing legislation: The Department of Agriculture Organic Act[18]*

The USDA's mission is to guide agriculture, nutrition, natural resources, and public policy based on agricultural data and management. The USDA supports farming, forestry, and ranching, all while leveraging satellite-provided data. In particular, the National Agricultural Statistics Service (NASS) can carry out its function largely due to space-based remote sensing data, as well as the analyses performed by Forest and Rangeland Research.

| Line Item | FY25 | FY24 | FY23 |
|---|---|---|---|
| **National Agricultural Statistics Service (NASS)** | 196 [19] | 187.513 [20] | 211 [*, 21] |
| **Forest and Rangeland Research** | 315.6 [22] | 300. [23] | 307 [†, 24] |
| Civil Space Estimate: | **511.6** | **487.513** | **518** |

*AMOUNTS IN MILLIONS.*

---

[*] Actual spending.
[†] Actual spending.

### Department of Commerce (DOC)

*Appropriations bill: Commerce, Justice, and Science*

*Establishing legislation: S.359—An Act to Establish the Department of Commerce and Labor[25]*

The primary mission of the DOC is to enhance and support job creation and economic growth in the United States by supporting commercial activity. The DOC, while supporting other significant space-related agencies such as NOAA and the Office of Space Commerce (OSC) (whose entries appear later in this section), also conducts space-related activities that affect trade, the economy, and the space workforce. For example, the Bureau of Economic Analysis (BEA) has recently added space as one of the sectors of the U.S. economy on which it publishes annual statistics.[26]

The DOC also advocates for the U.S. space industry. For example, the International Trade Admin-istration (ITA) "promotes U.S. aerospace companies exporting to international markets."[27] In particular, the ITA advocacy center, within its Global Markets division (which enforces export controls on commercial space), seeks to advance U.S. leadership in the global space economy, and in FY23 managed 27 cases related to "launch services, satellite manufacturing, satellite telecommunications services, sensors, ground station equipment, and related goods and services" with a total value of approximately $8.8 billion.[28]

Similarly, the Export Administration division within the Bureau of Industry and Security included commercial and civil space in its report to Congress by continuing efforts on a "multi-year industrial assessment to evaluate the current health and competitiveness of the civil segment of the U.S. space industrial base."

| Line Item | FY25 [*] | FY24 [†] | FY23 [‡] |
|---|---|---|---|
| **The International Trade Administration (ITA)** | | | |
| Global Markets[29] | 387.539 | 384.714 | 373.132 |
| Bureau of Economic Analysis[30] | 52.262 | 51.61 | 50.303 |
| **Bureau of Industry and Security** | | | |
| Export Administration[31] | Not Listed | 81.043 | 76.028 |
| Civil Space Estimate: | **439.801** | **517.367** | **499.463** |
| *AMOUNTS IN MILLIONS.* | | | |

### Department of Energy (DOE)

*Appropriations bill: Energy and Water Development and Related Agencies*
*Establishing legislation: The Department of Energy Organization Act[32]*

The primary mission of the DOE is to enhance the energy security of the United States through technological innovation and steward-responsible nuclear management. In particular, the DOE Office of Science takes advantage of the fact that stars, accretion disks, and black holes probe extreme cases of matter in the universe and provide excellent test laboratories for fusion and fission. Consequently, the DOE provides Earth-based research on astrophysical phenomena, and in turn uses space-based science to improve life on Earth. In order to conduct research, the DOE also expands fundamental computing capabilities and our understanding of basic physics that affects everyday life. The DOE also ensures such knowledge benefits society in workforce development, offers safeguards and security, and responsibly provides safe and updated infrastructure for its facilities.

| Line Item | FY25 [33] | FY24 [34] | FY23 *, [35] |
|---|---|---|---|
| **Office of Science (Total)** | 8,583 | 8,240 | 8,100 |
| Accelerator R&D and Production | 31 | 29 | 27.436 |
| Workforce Development for Teachers and Scientists | 43 | 40 | 42 |
| Isotope R&D and Production | 184 | 130.193 | 109.451 |
| Safeguards and Security | 195 | 190 | 184.099 |
| Science Laboratories Infrastructure | 295 | 208.907 | 280.7 |
| Program Direction | 246 | 226.831 | 211.211 |
| Fusion Energy Sciences | 844.5 | 790 | 763.222 |
| Nuclear Physics | 833 | 804 | 805.196 |
| Advanced Scientific Computing Research | 1,153 | 1,016 | 1,068 |
| High Energy Physics | 1,231 | 1,200 | 1,166 |
| Basic Energy Sciences | 2,582 | 2,606.25 | 2,534 |
| **Civil Space Estimate:** | **7,637.5** | **7,241.181** | **7,191.315** |

*AMOUNTS IN MILLIONS.*

---

* Unless otherwise noted, values are enacted.

### Department of Homeland Security (DHS)

*Appropriations bill: Department of Homeland Security*
*Establishing legislation: The Department of Homeland Security Act[36]*

The primary objective of the Department of Homeland Security is to protect the United States domestically through "customs, border, immigration enforcement, emergency response to natural and manmade disasters, antiterrorism work, and cybersecurity."[37] Within the DHS is the Cybersecurity and Infrastructure Security Agency (CISA), which protects U.S. space infrastructure resilience by monitoring space weather and space-based positioning, navigation, and timing (PNT).

Additionally, the *Air Security* project within Air, Land, and Port of Entry (POE) bridges capabilities between land, air, and space by enabling commercial satellite-based communications. As the U.S. Maritime Transportation System (MTS) has grown to include commercial space launches, its budget has expanded.[38] Additionally, the DHS FY25 report to Congress suggests DHS enforcement agencies seek to expand legacy communications systems in the Arctic with the capability and capacity to leverage commercial and civil space-based capabilities to "maritime regions devoid of infrastructure."[39]

| Line Item | FY25 | FY24 | FY23 |
|---|---|---|---|
| **CISA**[40] | 3,009 | 2,907 | 2,907 |
| **Border Security Thrust Area** | | | |
| Air, Land, and Port of Entry (POE) | | | |
| Air Security[41] | 14 | 14 | 14 |
| Maritime Safety and Security | | | |
| Port and Waterway Resiliency[42] | 0.5 | — | — |
| Remote Maritime Technologies[43] | 9.6 | 9.6 | 9.6 |
| **Civil Space Estimate:** | **3,033.1** | **2,930.6** | **2,930.6** |

*AMOUNTS IN MILLIONS.*

## Department of the Interior (DOI)

*Appropriations bill: Department of the Interior, Environment, and Related Agencies*
*Establishing legislation: An Act to Establish the Home Department, and to Provide for the Treasury Department an Assistant Secretary of the Treasury, and a Commissioner of the Customs[44]*

The guiding mission of the DOI is to manage domestic natural resources and safeguard cultural heritage. The DOI, like NOAA, is responsible for large contributions in financing, planning, and managing national and global remote sensing capabilities. The United States Geological Survey (USGS) conducts science for imaging land development, including the purchase of—and support to—commercial data services, and main-tains the Landsat civil remote sensing satellite program.[45] The next Landsat mission, *Landsat Next,* is set to launch in late 2030 and will be jointly operated with NASA to continue the world's longest space-based record of Earth's land surface. Additionally, the National Park Service contains the Natural Sounds and Night Skies Division within the Natural Resource Stewardship and Science Directorate. This division's unique task is to preserve natural darkness in the parks and allow park guests to experience the stars.

Of note in the FY25 budget is the start of a commercial satellite data pilot project. Other line items, such as *National Land Use Data Products*, are the analysis and interpretation of data acquired by remote sensing.

| Line Item | FY25 [46] | FY24 [47] | FY23 *, [48] |
|---|---|---|---|
| **Reported USGS Core Science Systems (Total)** | 313.016 [49] | 273.221 | 284.607 |
| Reported USGS National Land Imaging Program (Total) | 143.8 [xxii] | 115.071 | 115.921 |
| Reported Satellite Operations (Total) | 110.507 [50] | 95.334 | 92.184 |
| Sustainable Land Imaging Development - Landsat Next | 103.334 | 91.334 [51] | 0 [52] |
| Commercial Satellite Data Pilot | 5 | 0 [53] | 0 [54] |
| Baseline Capacity - 2024 Fixed Costs | 1.047 | 0[55] | 0 [56] |
| Reported Science Research and Investigations (Total) | 33.293 | 23.737 [57] | 0 [58] |
| Remote Sensing State Grants | 1.250 | 1.465 [59] | 0 [60] |
| Enhancing Landscape Measurements, Data, and Analysis | 3.7 | 0 [61] | 0 [62] |
| National Land Use Data Products | 1.5 | 0 [63] | 0 [64] |
| Natural Capital Accounting | 3.22 | 0.22 [65] | 0 [66] |
| Baseline Capacity - 2024 Fixed Costs | 1.036 | 0 [67] | 0 [68] |
| **National Park Service** | | | |
| Park Management | | | |
| Resource Stewardship | | | |
| Natural Resource Stewardship[69] | 273 | 272 | 250 † |
| Civil Space Estimate: | 393.087 | 365.019 | 250 |

*AMOUNTS IN MILLIONS.*

---

* Enacted amounts.
† Actual spending.

### Department of Labor (DOL)

*Appropriations bill: Departments of Labor, Health and Human Services, and Education*
*Establishing legislation: The Act of March 4, 1913 and 29 USC Ch. 12* [70]

Overall, the mission of the DOL is to promote the welfare of U.S. wage earners and to support job seekers and retirees. The DOL supports space activities through establishing terrestrial manufacturing standards and by publishing statistics on the health of the space economy. For example, the Occupational Safety and Health Administration (OSHA) sets standards and conducts workplace investigations, which are critical for minimizing the risk of potentially dangerous activities such as moving suspended heavy objects, transporting and loading rocket fuel and cryogenics, handling radioactive material, and handling large, pressurized cannisters.[71,72,73] OSHA's predicted contribution to the overall civil space budget is small; however, an exact estimate of its statistical relevance is incomplete as some of the largest space manufacturers have a six-year reporting gap to OSHA.[74] The health and growth of the overall workforce can be estimated from the Bureau of Labor Statistics studies and maintenance of the space commercial enterprise.[75]

| Line Item | FY25 | FY24 | FY23 [*] |
|---|---|---|---|
| **OSHA**[76] | 655.5 | 632.3 | 632.3 |
| **Bureau of Labor Statistics**[77] | 321 | 314 | 312 |
| Civil Space Estimate: | **976.5** | **946.3** | **944.3** |

*AMOUNTS IN MILLIONS.*

---

[*] Actual spending reported.

## Department of State (DOS)

*Appropriations bill: State, Foreign Operations, and Related Programs*
*Establishing legislation: An Act for Establishing an Executive Department, to be Denominated the Department of Foreign Affairs[78]*

The DOS guides U.S. foreign policy such that the interests and security of Americans are represented abroad. Related to space, the DOS represents the United States in bilateral space relationships with other countries and in multilateral institutions such as the United Nations. The DOS enforces international trade compliance and technology controls upstream in all areas of the space supply chain by ensuring joint missions with other countries comply with international information and technology sharing laws. The DOS secures downstream sharing via the Bureau of Political-Military Affairs (PM). The Directorate of Defense Trade Controls (DDTC) within the PM verify that space technology, craft components, and cargo with defense applications abide by International Traffic in Arms Regulations (ITAR).

| Line Item | FY25 [79] | FY24 [*, 80] | FY23 [†, 81] |
|---|---|---|---|
| **State Oceans and International Environmental and Scientific Affairs (OES)‡** | 21.949 | 17.192 [§, 82] | 15.192 |
| **Bureau of Arms Control, Deterrence, and Stability (ADS)** | 17.783 | 14.961 | 14.961 |
| **Bureau of Political-Military Affairs (PM)** | 15.9 | 13.6 | 10.6 |
| Civil Space Estimate: | **55.632** | **45.753** | **40.753** |

*AMOUNTS IN MILLIONS.*

---

[*] Following line items do not appear explicitly in *Consolidated Appropriations Act H.R. 4366/Public Law 118–42 Book 1 of 2 Divisions A–F*. www.govinfo.gov/content/pkg/CPRT-118HPRT56550/pdf/CPRT-118HPRT56550.pdf.
[†] Actual spending reported. Line items are not listed explicitly in *Consolidated Appropriations Act H.R. 4366/Public Law 118–42 Book 1 of 2 Divisions A–F*. www.govinfo.gov/content/pkg/CPRT-118HPRT56550/pdf/CPRT-118HPRT56550.pdf.
[‡] Includes the Office of Space Affairs.
[§] Does not appear explicitly in *Consolidated Appropriations Act H.R. 4366/Public Law 118–42 Book 1 of 2 Divisions A–F*. www.govinfo.gov/content/pkg/CPRT-118HPRT56550/pdf/CPRT-118HPRT56550.pdf.

### Department of Transportation (DOT)

*Appropriations bill: Transportation, Housing, and Urban Development*
*Establishing legislation: The Department of Transportation Act[83]*

The DOT is responsible for ensuring Americans have safe, efficient, reliable transportation within the country. In terms of civil space, the DOT provides oversight of space activities on the ground and as they transit the national airspace during launch and reentry.

On the ground, the Pipeline and Hazardous Materials Safety Administration regulates the transportation of hazardous materials to space manufacturing sites and the finished products (which are often large or hazardous) to launch sites.[84]

The FAA's Office of Commercial Space (FAA/AST) protects public safety during launch and reentry and prevents collisions between air and space traffic. To this end, the FAA Office of Policy, International Affairs, and Environment coordinates national airway closures during commercial space launches. Funding for the Commercial Space Transportation Safety Program has increased over time despite the encompassing budget for Research Engineering and Development decreasing.[85] These last two activities have expanded much in the last few years to accommodate the growing number of commercial space launch and reentry activities. According to the DOT budget request,[86] FY23 saw a "record-breaking number of requests for new authorizations." License modifications also broke a record in FY23 at 40 requests that fiscal year, which "exceeded the prior record by over 20%." AST also conducted 752 safety inspections in FY23 that resulted in another record-breaking number of operator noncompliance incidents and operator procedural errors, as well as seven concurrent mishap investigations.[87] AST anticipates the number of new authorizations will continue to increase, as many more commercial space launch vehicles are being developed.

Also due to the increased number of commercial space launches, the Air Traffic Organization (ATO) has added a new line item in 2025 for commercial space operations, and the budgetary increase to NextGen, the modernized air traffic management system, includes Enterprise Operational Performance Analysis to accommodate commercial space transportation.

| Line Item | FY25 | FY24 [88] | FY23 [*, 89] |
|---|---|---|---|
| **Federal Aviation Administration** | | | |
| Office of Chief Counsel | 74.03 | 55.78 [†, 90] | 55.78 [‡, 91] |
| Office of Policy, International Affairs, and Environment | 104.5 [92] | 90.4 [93] | 90.4 [94] |
| Wide Area Augmentation System (WAAS) for GPS | 91.8 | 92.1 | 73.2 |
| Research, Engineering, and Development (Total) | 250 [§,95] | 280 [**,96] | 255 [††, 97] |
| Commercial Space Transportation Safety Program | 5.4 [98] | 2 [99] | 4.708 [100] |
| NAS | | | |
| Air Traffic Control Facilities and Equipment | | | |
| Commercial Space Integration | 4.5 [101] | 1 | 5 [‡‡, 102] |
| Office of Commercial Space Transportation (AST) Operations | 57.1 [103] | 42.018 | 37.854 |
| Next Generation Air Transportation System (NextGen) | 73.556 | 65.581 [104] | 65.581 |
| Facilities and Equipment | | | |
| Advanced Technology Development and Prototyping[§§] | 31.9 | 31.9125 [105] | 29.45 |
| Air Traffic Organization (ATO) | | | |
| Commercial Space Operations | 21 | Not listed | Not listed |
| **Pipeline and Hazardous Materials Safety Administration** | 600.624 [***,106] | 519.382 [†††, 107] | 519.382 [‡‡‡, 108] |
| **PNT, GNSS, GPS[109]** | 3 | 20 | 20 |
| Civil Space Estimate:[§§§] | **1,312.01** | **1,198.174** | **1,151.643** |

*AMOUNTS IN MILLIONS.*

### Federal Communications Commission (FCC)

*Appropriations bill: Financial Services and General Government, previously CJS [110]*
*Establishing legislation: The Communications Act of 1934[111]*
*Department: N/A (Independent Agency)*

The Federal Communications Commission (FCC) is an independent regulatory agency with oversight of telecommunications. The FCC Space Bureau, established in 2023, is responsible for "developing, recommending, and administering policies, rules, standards, and procedures for the authorization and regulation of domestic and international satellite systems."[112] While mainly focused on use of radio frequency spectrum by satellites and ground stations, the FCC Space Bureau also has a requirement to protect the public interest and therefore has also created rules to mitigate risks from space debris, satellite collisions, and atmospheric reentry of the satellites it licenses.[113] To complement this work, the Office of International

Affairs represents the spectrum interests of the United States in international fora.[114]

Furthermore, the FCC Space Bureau organizes coordination agreements with the NSF and satellite licensees to mitigate damaging effects of unintentional light and radio interference with astronomical observatories managed by the NSF, which has led to cutting-edge innovations in satellite technology such as precise radio beaming[115] and light shielding.[116]

The Office of Engineering and Technology within the FCC has been developing new space communications abilities, "conducting engineering and technical analyses in advanced phases of terrestrial and space communications and special projects to obtain theoretical and experimental data on new or improved spectrum access and sharing techniques."[117]

| Line Item | FY25 | FY24 | FY23 * |
|---|---|---|---|
| **Office of Engineering and Technology** | 16.650 [118] | 15.600 [119] | 14.501 [120] |
| **Space Bureau** | 12.172 [121] | 11.125 [122] | 4.896 [123] |
| **Office of International Affairs (OIA)** | 10.152[124] | 9.594 [125] | 4.620 [126] |
| Civil Space Estimate: | **38.974** | **36.319** | **24.017** |

*AMOUNTS IN MILLIONS.*

---

* Actual spending.

### National Aeronautics and Space Administration (NASA)

***Appropriations bill: Commerce, Justice, and Science***
***Establishing legislation: The National Aeronautics and Space Act of 1958[127]***
***Department: N/A (Independent Agency)***

The primary mission of NASA is space-based, which is to innovate technology that enables space exploration and expands human knowledge of the cosmos. As a result, NASA contains the largest number of line items related to civil space and is worthy of its own dedicated discussion beyond the overview scope of this paper. Detailed studies have been conducted,[128] as well as one that includes recent supplementary funding outside the appropriations bill.[129] Where available for individual line items, supplementary funding has been included in the table below to provide a more realistic sense of historic spending. As the scope of this paper is the civil space budget, the table on the next page does not include the NASA funding for the aeronautics part of its mission.

Many line items for NASA have changed names over the last few years. For example, the *Space Launch System* (SLS) was originally under *Exploration/Deep Space Exploration Systems*, but in FY25 its budget is a part of the *Moon to Mars Transportation System*. Additionally, many line items related to the Artemis Mission are found under *Moon to Mars* program headers. The *Artemis Mission* contains several phases for each technological benchmark. The program has spread across multiple appropriations cycles but appears under the same name, *Artemis Mission*. For additional details and an updated status of the mission, please visit the Artemis section of: https://www.nasa.gov/nasa-missions/.

*Mission Services and Capabilities* (MSaC) supports all NASA projects through overhead responsibilities, such as financial management, technology services, legal assistance, procurement, and small business initiatives. Additionally, technology maturation and small business innovation research are included in the *Space Technology (Total)* budget line.[130]

As civil space has matured, the NASA budget also includes commercial LEO development partnerships, dedicated to incorporating commercial innovation and developing a "robust commercial space economy" while also ensuring NASA can meet its mission needs. A primary example of this is the *Commercial LEO Development* program to support privately managed space stations after the International Space Station is decommissioned.[131] Similarly, *Space Transportation* budget line funds the development of commercial capabilities for the safe, reliable, and affordable transport of astronauts and cargo to and from LEO, as well as the deorbiting of the ISS.[132] *Deep Space Exploration* empowers commercial capabilities to explore the lunar surface and create habitats.[133]

| Line Item | FY25 [134] | FY24 [*, 135] | FY23 [†, 136] |
|---|---|---|---|
| **Deep Space Exploration Systems (Total)** | 7,618.2 | 7,666 [‡,137] | 7,468.850 |
| Moon to Mars Transportation System | 4,213.0 | 4,533 | 4,738 |
| Moon to Mars Lunar Systems Development | 3,288.1 | Not listed | 2,600 |
| Human Exploration Requirements and Architecture | 117.1 | Not listed | 100.5 |
| **Space Operations (Total)** | 4,389.7 | 4,220 [§, 138] | 4,250 |
| International Space Station | 1,269.6 | Not listed | 1,286.2 [139] |
| Space Transportation | 1,862.1 | Not listed | 1,759.6 [140] |
| Space and Flight Support (SFS) | 1,088.4 | Not listed | 983.4 [141] |
| Commercial LEO Development | 169.6 | 228 | 224.3 |
| Exploration Operations | — | Not listed | 13.2 |
| **Space Technology (Total)** | 1,181.8 | 1,100 | 1,193 |
| **Science (Total)** | 7,565.7 | 7,334 | 7,791.5 [142] |
| Earth Science | 2,378.7 | 2,195 | 2,195 |
| Planetary Science | 2,731.5 | 2717 | 3,216.5 [143] |
| Astrophysics | 1,578.1 | 1530 | 1,510.0 |
| Heliophysics | 1,578.1 | 805 | 805.0 |
| Biological and Physical Sciences | 90.8 | 88 | 85.0 |
| **STEM Engagement (Total)** | 143 | 143 | 143.5 [144] |
| **Safety, Security, and Mission Services (Total)** | 3,044.4 | 3,129 | 3,129 |
| Mission Services and Capabilities | 2,058.1 | Not listed | 2,067.4 [145] |
| Engineering, Safety, and Operations | 986.3 | Not listed | 1,069.1 [146] |
| **Construction and Environmental Compliance and Restoration (Total)** | 424.1 | 300 | 604 |
| Construction of Facilities | 344.7 | — | 346.2 [147] |
| Environmental Compliance and Restoration | 79.4 | — | 76.2 [148] |
| **Inspector General (Total)** | 50.5 | 48 | 47.6 |
| NASA Reported Total [**]: | **25,383.7** | **24,875** | **25,573** [††] |

*AMOUNTS IN MILLIONS.*

[*] Enacted amounts.
[†] Enacted amounts.
[‡] According to Lindbergh 2024, "Includes 3.133B appropriated to Exploration account by PL 118-42 without specified purpose."
[§] According to Lindbergh 2024, "Includes 2.136B appropriated to space operations account by PL 118-42 without specified purpose."
[**] Includes Aeronautics, not included in this civil space study.
[††] According to Lindbergh 2024, "Includes 556.4M in emergency supplemental funding provided in Division N of PL 117-328."

### National Institute of Standards and Technology (NIST)

*Appropriations bill: Commerce, Justice, and Science*
*Establishing legislation: An Act to establish the National Bureau of Standards[149]*
*Department: DOC*

NIST is a government agency with deep expertise in creating industry standards and standards of measurement to promote U.S. innovation and competitiveness. As commercial space develops and civil space capability expands, there is a growing need to establish technical standards and research for a space-based future. NIST's civil space functions include developing domestic standards and representing U.S. interests in international standards bodies. NIST staff also directly support the work of the Office of Space Commerce to promote U.S. commercial space.

| Line Item | FY25 [150] | FY24 [151] | FY23 *, [152] |
|---|---|---|---|
| **STRS Scientific and Technical Research and Services** | 975 | 1,080 | 953 |
| Standards Coordination and Special Programs† | Not listed | Not listed | Not listed |
| Civil Space Estimate: | **975** | **1,080** | **953** |

*AMOUNTS IN MILLIONS.*

---

### National Oceanic and Atmospheric Administration (NOAA)

**Appropriations bill: Commerce, Justice, and Science**
**Establishing legislation: The Reorganization Plan No. 4 of 1970[153]**
**Department: DOC**

NOAA has the responsibility of researching, analyzing, and disseminating knowledge as it relates to weather, climate, oceans, and the atmosphere. NOAA primarily engages with civil space through its remote sensing projects in NESDIS (National Environmental Satellite, Data, and Information Service) and the Earth observations group, U.S. Group on Earth Observations (USGEO).

Additionally, NOAA utilizes space-based navigation and observation capabilities critical for operations in the ocean. NOAA also houses the Office of Space Commerce, which is developing a space situational awareness tool called the *Traffic Coordination System for Space* (TraCSS) and is providing oversight of commercial space activities, especially through its Commercial Remote Sensing Regulatory Affairs (CRSRA) office.

One recent change is the discontinuation of the *NESDIS Community Project*, which was a one-time congressionally directed project provided in the FY23-enacted bill.[154]

| Line Item | FY25 [155] | FY24 [156] | FY23 [157] |
|---|---|---|---|
| **NESDIS (Total)** | 397.51 | 380.765 | 375.537 [158] |
| NESDIS Environmental Satellite Observing Systems (Total) | 324 | 310.765 | 304 |
| NESDIS Office of Satellite and Product Operations | 262 | 250.165 | 245.9 |
| NESDIS Product Development, Readiness, and Application | 61 | 59.850 | 57.5 |
| NESDIS US Group on Earth Observations (USGEO) | 1 | 0.750 | 0.75 |
| NESDIS National Centers for Environmental Information (Total) | 73 | 70 | 71.4 |
| NESDIS NOAA Community Project Funding/NOAA Special Projects (Total) | 0 | 0 | 2.5 |
| **Mission Support** | | | |
| Office of Space Commerce | 75.638 [159] | 65 | 70 [160] |
| **National Ocean Service (NOS) Navigation, Observations, and Positioning** | | | |
| National Geodetic Survey (NGS)* | 207.7 [161] | 257.702 | 259.7 [162] |
| Civil Space Estimate: | **680.848** | **703.467** | **707.75** |

*AMOUNTS IN MILLIONS.*

---

* Maintains the Continuously Operating Reference Stations (CORS).

### National Science Foundation (NSF)

*Appropriations bill: Commerce, Justice, and Science*
*Establishing legislation: the National Science Foundation Act of 1950[163]*
*Department: N/A (Independent Agency)*

The National Science Foundation (NSF) issues grants to conduct scientific research. This includes both analyzing data of astrophysical phenomena and operating major observatories in the United States, as well as international research outposts such as in Antarctica and Chile. The impact of the NSF reaches all areas of science, and therefore includes significant space and space-adjacent activity, such as space weather studies through the National Geophysical Facility (NGF), with the "aim of increasing resilience to such natural hazards."[164] Additionally, NSF manages the *National Center for Atmospheric Research* (NCAR), which researches the Earth-Sun system, and the *National Optical-Infrared Astronomy Research Laboratory* (NOIRLab), which contains cutting-edge astronomical endeavors such as the *Gemini Observatory* and the upcoming *Vera C. Rubin Observatory*.

Much of the NSF mission area spans artificial intelligence, biotechnology, and microelectronics, not always related to astronomy, space, or physics. Within the NSF budget request, there were no major educational or STEM programs that focused solely on space, physics, or astronomy. As such, STEM-related investments and education have been omitted from this list.

The NSF additionally has several divisions under the Mathematical and Physical Sciences (MPS) Directorate that relate to space. In particular, astronomical sciences (AST) and physics (PHY) bear a direct relation to civil space. The table highlights several civil-space related projects within the NSF that receive funding from a combination of divisions that include AST and PHY that contribute to research, education, and infrastructure. The topline totals for each division are included for completeness but are not incorporated into the calculated total to avoid double-counting these contributions to specific projects. For a comprehensive breakdown of facilities and their contributions across all NSF divisions, see Summary Tables – 18 in the NSF Request to Congress.[165]

| Line Item | FY25 | FY24 [166] | FY23 |
|---|---|---|---|
| GEO Division AGS (Atmospheric and Geospace Sciences) Research, Education, and Infrastructure (Total) | 293.80 [167] | Not listed | 289.71[168] |
| MPS Division AST—Research, Education, and Infrastructure (Total) | 318.53[169] | Not listed | 288.21[170] |
| MPS Division PHY—Research, Education, and Infrastructure (Total) | 312.90[171] | Not listed | 308.65[172] |
| Advanced Wireless Funding* | 167.90 [173] | Not listed | 154.02[174] |
| Total Obligations for NGF (formerly GAGE and SAGE) | 45.29 [175] | Not listed | 37.92 [176] |
| Spectrum Innovation Initiative | 17 [177] | Not listed | 17 [178] |
| Green Bank Observatory | 9.68 [179] | Not listed | 10.83 [180] |
| National Center for Atmospheric Research (NCAR) | 124.59 [181] | Not listed | 116.20 [182] |
| IceCube Neutrino Observatory | 8.23 [183] | Not listed | 7.66 [184] |
| Laser Interferometer Gravitational-Wave Observatory | 49 [185] | Not listed | 45 [186] |
| NOIRLab | 86.40 [187] | Not listed | 73.57 [188] |
| National Radio Astronomy Observatory | 96.71 [189] | Not listed | 93.66 [190] |
| Antarctic Facilities and Operations | 269.94 [191] | Not listed | 224.71 [192] |
| National Solar Observatory | 34.24 [193] | Not listed | 26.56 [194] |
| Vera C. Rubin Observatory (Construction) | 7.61[195] | Not listed | 15 [196] |
| Other Research Infrastructure† | 131.62[197] | Not listed | 130.21[198] |
| **Civil Space Estimate:‡** | **1,048.21** | **Not listed** | **952.34** |

*AMOUNTS IN MILLIONS.*

\* Cross-cutting topic that includes MPS funding and no GEO funding.
† Includes Center for High Energy X-ray Science (CHEXS) and support for the Arecibo Observatory.
‡ Does not include MPS and GEO totals to avoid double-counting facilities and interdisciplinary topics.

**National Telecommunications and Information Administration (NTIA)**
*Appropriations bill: Commerce, Justice, and Science*
*Established by: Executive Order 12046[199]*
*Department: DOC*

The NTIA regulates spectrum for government space missions in tandem with the FCC, which regulates private sector spectrum use. The NTIA's advanced communications research builds on decades of research by the NTIA's labs in Boulder, CO that model and analyze spectrum occupancy and evaluate the Spectrum Access System and Environmental Sensing Capability (SAS/ESC) to better predict impacts for commercial providers. Both line items in the NTIA's budget faced a decrease in FY25, following a large increase from FY23 to FY24.

| Line Item | FY25 | FY24 [*] | FY23 [†,‡] |
|---|---|---|---|
| **Advanced Communications Research** | 14.4 [200] | 17.2 [201] | 11.6 [202] |
| **Spectrum Management** | 11 [203] | 16.9 [204] | 3.3 [205] |
| Civil Space Estimate: | **25.4** | **34.1** | **14.9** |

*AMOUNTS IN MILLIONS.*

---

[*] Following line items do not appear explicitly in *Consolidated Appropriations Act H.R. 4366/Public Law 118–42 Book 1 of 2 Divisions A–F*. www.govinfo.gov/content/pkg/CPRT-118HPRT56550/pdf/CPRT-118HPRT56550.pdf.
[†] Actual spending.
[‡] Following line items do not appear explicitly in *Consolidated Appropriations Act H.R. 4366/Public Law 118–42 Book 1 of 2 Divisions A–F*. www.govinfo.gov/content/pkg/CPRT-118HPRT56550/pdf/CPRT-118HPRT56550.pdf.

### The Nuclear Regulatory Commission (NRC)

*Appropriations bill: Energy and Water Development*
*Establishing legislation: The Energy Reorganization Act of 1974[206]*
*Department: N/A (Independent Agency)*

The NRC has the primary mission of both licensing and regulating radioactive materials under civilian use, in addition to protecting the health and safety of the public and environment as they relate to use of nuclear materials. It is also a member of the Interagency Nuclear Safety Review Board (INSRB), who evaluates the safety of government space missions that have nuclear sources onboard.[207] The NRC ensures that, with the civilian use of radioactive materials, there is still a protection of public health and safety and promotion of common defense and security. Most notably, the INSRB was responsible for a safety analysis of the Mars 2020 Perseverance rover, which had a radioisotope generator onboard.[208]

| Line Item | FY25 | FY24 | FY23 |
|---|---|---|---|
| **Nuclear Regulatory Commission (Total)[209]** | 974.946 | 927.153 | 881.466 * |

*AMOUNTS IN MILLIONS.*

---

\* Actual spending.

## The Smithsonian Institution

*Appropriations bill: Department of the Interior, Environment, and Related Agencies*
*Establishing legislation: An Act to Establish the "Smithsonian Institution for the Increase and Diffusion of Knowledge among Men"[210]*
*Department: Independent Trust Instrumentality of the United States*

The Smithsonian Institution was founded with the primary purposes of both increasing and disseminating knowledge. Related to civil space, the Smithsonian Institution's mission is to inspire visitors with our planet's uniqueness in the universe and to encourage future generations to seek careers in the space workforce. The mission of the Smithsonian Institution builds on scientific data from space missions and disseminates that data into publicly accessible language. In recent years, the *Smithsonian Astrophysical Observatory* and the *Air and Space Museum* have requested budget increases for important facilities upgrades and to support programming celebrating the history and growth of the U.S. space program and new commercial space activities, while the *Universe Center* has decreased in budget.

| Line Item | FY25 | FY24 [211] | FY23 *, [212] |
|---|---|---|---|
| **Major Scientific Instrumentation (Standalone)** | 4.1 [213] | 4.118 | 4.118 |
| **National Air and Space Museum** | 23.7 [214] | 22.380 | 22.380 |
| **Smithsonian Astrophysical Observatory (SAO)** | 27.743 [215] | 26.412 | 26.412 |
| **Facilities Capital** | | | |
| Smithsonian Astrophysical Observatory† | 5 [216] | Not listed | 1.8 [217] |
| Air and Space Museum‡ | 8 [218] | Not listed | 5.5 § |
| **Enhanced Research** | | | |
| Universe Center | 0.182 [219] | Not listed | 0.184 [220] |
| Civil Space Estimate: | **68.725** | **52.91** | **60.39** |

*AMOUNTS IN MILLIONS.*

# Civil Space by National Priority

The following section lists the same budget information as before but grouped by categories of national priorities that span multiple departments and agencies or include different parts of an agency's budget. Although some budget lines may qualify for more than one category, each is represented only once, based on the analyst's judgment of the category to which it contributes the most.

## American Leadership and Manufacturing

The United States has placed a high priority on expanding domestic manufacturing and creating secure and stable supply chains domestically and abroad. These efforts enable greater control over production processes and supply chains, ensure quality standards and reduce reliance on foreign sources, and contribute to national economic resilience and growth. Similarly, upholding strong safety and product standards (such as adequate training requirements) and ensuring safe working conditions through OSHA contributes to sustainable economic growth while lowering healthcare costs.

In particular, NASA's *Mission Services and Capabilities* (MSaC) offers a comprehensive suite of support services crucial to enabling challenging and complex space operations. Similarly, NASA's Deep Space Exploration funds the domestic manufacturing of lunar systems and physical architecture for space exploration.

American Leadership and Manufacturing also includes efforts to bolster small businesses and domestic suppliers throughout the entire value chain while protecting national security through export controls on critical defense technologies. U.S. manufacturing interests are represented abroad in international fora by several agencies, such as the DOS, FCC, and NTIA, which helps ensure U.S. leadership in international institutions that develop interoperability and safety standards, assign radio frequency spectrum allocations that are vital to many space capabilities, and ensure the long term sustainability of space activities. Federal investments in emerging technologies and incentives for their wider adoption enables the United States to maintain a competitive advantage in the global market; the development and deployment of these technologies is ushered by the ITA and FCC.

| Line Item | Agency | FY25 | FY24 | FY23 |
|---|---|---|---|---|
| **State Oceans and International Environmental and Scientific Affairs (OES)** | DOS | 21.949 | 17.192 | 15.192 |
| **OSHA** | DOL | 655.5 | 632.3 | 632.3 |
| **Safety, Security, and Mission Services (Total)** | NASA | 3,044.4 | 3,129 | 3,129 |
| Mission Services and Capabilities | NASA | 2,058.1 | Not listed | 2,067.4 |
| Engineering, Safety, and Operations | NASA | 986.3 | Not listed | 1069.1 |
| **Deep Space Exploration Systems (Total)** | NASA | 7,618.2 | 7,666 | 7,468.850 |
| Moon to Mars Transportation System | NASA | 4,213.0 | 4,533 | 4,738 |
| Moon to Mars Lunar Systems Development | NASA | 3,288.1 | Not listed | 2,600 |
| Human Exploration Requirements and Architecture | NASA | 117.1 | Not listed | 100.5 |
| **STRS Scientific and Technical Research and Services** | NIST | 975 | 1,080 | 953 |
| Standards Coordination and Special Programs | NIST | Not listed | Not listed | Not listed |
| **Bureau of Industry and Security** | DOC | | | |
| Export Administration | DOC | Not reported | 81.043 | 76.028 |
| **The International Trade Administration (ITA)** | DOC | | | |
| Global Markets | DOC | 387.539 | 384.714 | 373.13 |
| **Mission Support** | NOAA | | | |
| Office of Space Commerce | NOAA | 75.638 | 65 | 70 |
| **Office of Engineering and Technology** | FCC | 16.650 | 15.600 | 14.501 |
| **Office of International Affairs (OIA)** | FCC | 10.152 | 9.594 | 4.620 |
| **Advanced Communications Research** | NTIA | 14.4 | 17.2 | 11.6 |
| **Office of Science** | DOE | | | |
| Accelerator R&D and Production | DOE | 31 | 29 | 27.436 |
| Advanced Scientific Computing Research | DOE | 1,153 | 1,016 | 1,068 |
| Isotope R&D and Production | DOE | 184 | 130.193 | 109.451 |
| **Advanced Wireless Funding** | NSF | 167.90 | Not listed | 154.02 |
| **Space Operations** | NASA | | | |
| Commercial LEO Development | NASA | 169.6 | 228 | 224.3 |
| **Space Technology (Total)** | NASA | 1,181.8 | 1,100 | 1,193 |
| **Federal Aviation Administration** | DOT | | | |
| NAS | DOT | | | |
| Air Traffic Control Facilities and Equipment | DOT | | | |
| Commercial Space Integration | DOT | 4.5 | 1 | 5 |
| **Reported USGS Core Science Systems** | DOI | | | |
| Reported USGS National Land Imaging Program | DOI | | | |
| Commercial Satellite Data Pilot | DOI | 5 | 0 | 0 |
| Civil Space Estimate* | | **15,716.23** | **15,601.836** | **15,529.43** |

*AMOUNTS IN MILLIONS.*

---

* This estimate includes office total budgets and individual line items. Some lines were not reported for FY25.

## Workforce Development

Achieving preeminence in space activities necessitates a highly trained, skilled, and dedicated workforce. Cultivating this talent pool requires leveraging educational initiatives such as those championed by NASA and the Smithsonian Institution, which inspire and engage the nation's youth with science, technology, engineering, and mathematics (STEM) education. These STEM programs not only foster scientific literacy but also instill a sense of national purpose and ambition regarding space exploration.

The economic potential of space development relies on creating high-skilled, well-paying jobs that are accessible to the broader population across the country. Investment in this employment pipeline fosters a growing labor force that will contribute to the burgeoning space supply chain, participating directly in the exploration and utilization of this final frontier.

Quantifying the success and vitality of these endeavors requires ongoing monitoring and analysis. The Bureau of Labor Statistics provides valuable data on workforce trends, while the National Science Foundation plays a crucial role in fostering education through programs like Spectrum Innovation. These initiatives focus on equipping the workforce with the skills necessary to navigate the evolving technological landscape, including spectrum management and sharing techniques, as demand for radio frequencies intensifies across various sectors.

| Line Item | Agency | FY25 | FY24 | FY23 |
|-----------|--------|------|------|------|
| **Bureau of Labor Statistics** | DOL | 321 | 314 | 312 |
| **STEM Engagement (Total)** | NASA | 143 | 143 | 143.5 |
| **National Air and Space Museum** | Smithsonian | 23.7 | 22.380 | 22.380 |
| **Facilities Capital** | Smithsonian | | | |
| Air and Space Museum | Smithsonian | 8 | Not listed | 5.5 |
| **Office of Science** | DOE | | | |
| Workforce Development for Teachers and Scientists | DOE | 43 | 40 | 42 |
| **Spectrum Innovation Initiative** | NSF | 17 | Not listed | 17 |
| **National Park Service** | DOI | | | |
| Park Management | DOI | | | |
| Resource Stewardship | DOI | | | |
| Natural Resource Stewardship | DOI | 273 | 272 | 250 |
| Civil Space Estimate: | | **828.7** | **791.38** | **792.38** |

*AMOUNTS IN MILLIONS.*

## Fundamental Science

Establishing a long track record of major scientific breakthroughs that lead to successful commercial applications and societal benefits begins with sustained, long-term investments in fundamental scientific research. As this baseline grows, so does the potential for innovative applications of scientific advancement that make companies throughout the United States more competitive and, at the same time, increase their ability to deliver new and improved benefits to the American people.

A large portion of the most important space research happens on Earth. For decades, the desire to travel to remote destinations in the solar system was fueled by research conducted in terrestrial observatories through funding from NASA and the NSF and disseminated by the Smithsonian Institution. Today, new advancements in nuclear propulsions to make those dreams manifest are supported by the DOE.

| Line Item | Agency | FY25 | FY24 | FY23 |
|---|---|---|---|---|
| **Major Scientific Instrumentation (Standalone)** | Smithsonian | 4.1 | 4.118 | 4.118 |
| **Smithsonian Astrophysical Observatory (SAO)** | Smithsonian | 27.743 | 26.412 | 26.412 |
| **Facilities Capital** | Smithsonian | | | |
| Smithsonian Astrophysical Observatory | Smithsonian | 5 | Not listed | 1.8 |
| **Enhanced Research** | Smithsonian | | | |
| Universe Center | Smithsonian | 0.182 | Not listed | 0.184 |
| **Office of Science** | DOE | | | |
| High Energy Physics | DOE | 1,231 | 1,200 | 1,166 |
| Basic Energy Sciences | DOE | 2,582 | 2,606.25 | 2,534 |
| **IceCube Neutrino Observatory Funding** | NSF | 8.23 | Not listed | 7.66 |
| **Laser Interferometer Gravitational-Wave Observatory Funding** | NSF | 49 | Not listed | 45 |
| **Vera C. Rubin Observatory (Construction)** | NSF | 7.61 | Not listed | 15 |
| **Green Bank Observatory** | NSF | 9.68 | Not listed | 10.83 |
| **NOIRLab** | NSF | 86.40 | Not listed | 73.57 |
| **Other Research Infrastructure** | NSF | 131.62 | Not listed | 130.21 |
| **Science** | NASA | | | |
| Planetary Science | NASA | 2,731.5 | 2,717 | 3,216.5 |
| Astrophysics | NASA | 1,578.1 | 1530 | 1,510 |
| Biological and Physical Sciences | NASA | 90.8 | 88 | 85.0 |
| **Space Operations** | NASA | | | |
| International Space Station | NASA | 1,269.6 | Not listed | 1,286.2 |
| Exploration Operations | NASA | — | Not listed | 13.2 |
| Civil Space Estimate[*]: | | **9,812.565** | **8,171** | **10,125.684** |

*AMOUNTS IN MILLIONS.*

---

## Efficiency, Improvements, and Growth

Government efficiency hinges on the timely implementation of programs authorized and funded by congressional appropriations. In this manner, fiscal responsibility takes on a comprehensive, enterprise-wide perspective that accounts for the costs of *ending* programs as well as continuing them. With a holistic view of government operations (performed in the executive office of the president and informed by the Bureau of Economic Analysis), calculated adjustments can be implemented with minimal disruption to overall operations, refining processes and improving outcomes without drastic overhauls that would risk destabilizing essential government functions and require expensive reversals.

Improvements are carried out through program directors and chief counsel at different agencies within the space sector, such as the DOT and DOE, and growth into new sectors moves forward by adopting new technologies[*] and facilitating commercial space operations within the DOT.

| Line Item | Agency | FY25 | FY24 | FY23 |
|---|---|---|---|---|
| **Office of Management and Budget (OMB)** | Executive Office of the President | 138.3 | 128.035 | 128.035 |
| **Bureau of Economic Analysis** | DOC | 52.262 | 51.61 | 50.303 |
| **National Space Council** | Executive Office of the President | 2 | 1.965 | 1.965 |
| **Office of Science and Technology Policy (OSTP)** | Executive Office of the President | 8 | 7.965 | 7.965 |
| **Federal Aviation Administration** | DOT | | | |
| Next Generation Air Transportation System (NextGen) | DOT | 73.556 | 65.581 | 65.581 |
| Office of Chief Counsel | DOT | 74.03 | 55.78 | 55.78 |
| Office of Policy, International Affairs, and Environment | DOT | 104.5 | 90.4 | 90.4 |
| Air Traffic Organization (ATO) | DOT | | | |
| Commercial Space Operations | DOT | 21 | Not listed | Not listed |
| **Office of Science** | DOE | | | |
| Program Direction | DOE | 246 | 226.83 | 211.211 |
| Civil Space Estimate: | | **719.648** | **628.167** | **611.24** |

*AMOUNTS IN MILLIONS.*

[*] New technologies, by nature, can introduce uncertainties for long-term planning and policy. For a discussion of cutting-edge technology adaptation and long-term support, see https://csps.aerospace.org/papers/strategic-foresight-addressing-uncertainty-long-term-strategic-planning.

### Homeland Security

Homeland security encompasses safeguarding vital infrastructure, ensuring public safety, and mitigating risks posed by various actors and events. While the Department of Defense (DOD) plays a significant role in national defense, homeland security prioritizes functions that are important for national security but are beyond the scope of traditional military operations. This includes the protection of critical civilian assets and applications from dual-use systems such as GPS, cybersecurity, and disaster response.

The greatest financial contribution to this category is the DHS agency Cybersecurity and Infrastructure Security Agency (CISA), which focuses on protecting U.S. dams, reservoirs, factories, emergency services, and communications infrastructure from cyber and physical threats, including those in space that monitor weather and navigation. In addition to contributions by the DHS, homeland security as a category also benefits from the actions of the State Department. The State Department similarly takes on roles that liaise between the United States and foreign actors through means of commerce and international cooperation.

Furthermore, NSF initiatives like the *Global Aerospace Geodetic Experiment* (GAGE) and the *Seismic Array for Global Earth Observation* (SAGE), which monitor GPS (along with NOAA's *National Geodetic Survey* (NGS)) and seismic activity respectively, provide a scaffolding of scientific data to mitigate potential threats from human actors and risks posed by natural disasters.[221]

| Line Item | Agency | FY25 | FY24 | FY23 |
|---|---|---|---|---|
| **CISA** | DHS | 3,009 | 2,907 | 2,907 |
| **Bureau of Arms Control, Deterrence, and Stability (ADS)** | State | 17.783 | 14.961 | 14.961 |
| **Bureau of Political-Military Affairs (PM)** | State | 15.9 | 13.6 | 10.6 |
| **National Ocean Service (NOS) Navigation, Observations, and Positioning** | NOAA | | | |
| National Geodetic Survey (NGS) | NOAA | 207.7 | 257.702 | 259.7 |
| **Total Obligations for NGF (formerly GAGE and SAGE)** | NSF | 45.29 | Not listed | 37.92 |
| **PNT, GNSS, and GPS** | DOT | 3 | 20 | 20 |
| **Wide Area Augmentation System for GPS** | DOT | 91.8 | 92.1 | 73.2 |
| Civil Space Estimate: | | **3,390.47** | **3,305.363** | **3,323.401** |

*AMOUNTS IN MILLIONS.*

### Infrastructure, Energy, and Resiliency

Sending astronauts, specialized equipment, and robotic spacecraft into the harsh space environment requires planning, oversight, and cooperation from diverse sectors on the ground. Multiple agencies and departments contribute to a reliable infrastructure for civil space, ensure the energy is available to power each stage of the process, and maintain a resilient presence in space. Infrastructure itself includes launch, communications, and space propulsion, and power systems.

Getting a payload into space and keeping it operational requires infrastructure on the ground and in orbit. Movement of components, assemblies, and finished products through the entire manufacturing process requires careful transportation along waterways, airways, and roadways. Once in space, the ability to predict and prepare for solar weather events is essential to operational resilience. These natural phenomena are monitored and studied by the NASA *Heliophysics* division and the NSF *National Solar Observatory.*

| Line Item | Agency | FY25 | FY24 | FY23 |
|---|---|---|---|---|
| **Border Security Thrust Area** | DHS | | | |
| Maritime Safety and Security | DHS | | | |
| Remote Maritime Technologies | DHS | 9.6 | 9.6 | 9.6 |
| Port and Waterway Resiliency | DHS | 0.5 | — | — |
| Air, Land, and Port of Entry (POE) | DHS | | | |
| Air Security | DHS | 14 | 14 | 14 |
| **Nuclear Regulatory Commission (Total)** | NRC | 974.946 | 927.153 | 881.466 |
| **Space Bureau** | FCC | 12.172 | 11.125 | 4.896 |
| **Spectrum Management** | NTIA | 11 | 16.9 | 3.3 |
| **Office of Science** | DOE | | | |
| Fusion Energy Sciences | DOE | 844.5 | 790 | 763.222 |
| Nuclear Physics | DOE | 833 | 804 | 805.196 |
| Science Laboratories Infrastructure | DOE | 295 | 208.907 | 280.7 |
| **National Solar Observatory** | NSF | 34.24 | Not listed | 26.56 |
| **Antarctic Facilities and Operations** | NSF | 269.94 | Not listed | 224.71 |
| **National Radio Astronomy Observatory** | NSF | 96.71 | Not listed | 93.66 |
| **Science** | NASA | | | |
| Heliophysics | NASA | 1,578.1 | 805 | 805.0 |
| **Space Operations** | NASA | | | |
| Space Transportation | NASA | 1,862.1 | Not listed | 1,759.6 |
| Space and Flight Support (SFS) | NASA | 1,088.4 | Not listed | 983.4 |
| **Construction and Environmental Compliance and Restoration (Total)** | NASA | 424.1 | 300 | 604 |
| Construction of Facilities | NASA | 344.7 | — | 346.2 |
| Environmental Compliance and Restoration | NASA | 79.4 | — | 76.2 |
| **Safeguards and Security** | DOE | 195 | 190 | 184.099 |
| **Pipeline and Hazardous Materials Safety Administration** | DOT | 600.624 | 519.382 | 519.382 |
| **Federal Aviation Administration** | DOT | | | |
| Facilities and Equipment | DOT | | | |
| Advanced Technology Development and Prototyping | DOT | 31.9 | 31.9125 | 29.45 |
| Research, Engineering, and Development (Total) | DOT | 250 | 280 | 255 |
| Commercial Space Transportation Safety Program | DOT | 5.4 | 2 | 4.708 |
| AST Operations | DOT | 57.1 | 42.018 | 37.854 |
| Civil Space Estimate: | | **9,912.43** | **4,951.998** | **8,712.203** |

*AMOUNTS IN MILLIONS.*

### *Remote Sensing Applications*

Remote sensing is more than funding the launch and operation of satellites. Just as important, or even more so, are the programs that fund the application of remote sensing data, which protect and enhance a growing range of other sectors such as forestry, agriculture, transportation, and weather forecasting. In particular, the growing demand for remote sensing data allows for private companies to offer services in this emerging market, offering data analyses and interpretations that further allow for better protecting the nation's natural resources and critical infrastructure, such as water and air quality, as well as defending our resources from methane leaks, wildfires, and other threats. Access to high-resolution, fast-turnaround remote sensing data saves valuable time and money, and pinpoints geographical locations of conflicts or damage.

| Line Item | Agency | FY25 | FY24 | FY23 |
|---|---|---|---|---|
| **NESDIS (Total)** | NOAA | 397.51 | 380.765 | 375.537 |
| NESDIS Environmental Satellite Observing Systems (Total) | NOAA | 324 | 310.765 | 304 |
| NESDIS Office of Satellite and Product Operations | NOAA | 262 | 250.165 | 245.9 |
| NESDIS Product Development, Readiness, and Application | NOAA | 61 | 59.850 | 57.5 |
| NESDIS US Group on Earth Observations (USGEO) | NOAA | 1 | 0.750 | 0.75 |
| NESDIS National Centers for Environmental Information (Total) | NOAA | 73 | 70 | 71.4 |
| NESDIS NOAA Community Project Funding/NOAA Special Projects (Total) | NOAA | 0 | 0 | 2.5 |
| **National Agricultural Statistics Service (NASS)** | USDA | 196 | 187.513 | 211 |
| **Forest and Rangeland Research** | USDA | 315.6 | 300 | 307 |
| **Science** | NASA | | | |
| Earth Science | NASA | 2,378.7 | 2,195 | 2,195 |
| **Reported USGS Core Science Systems** | DOI | | | |
| Reported USGS National Land Imaging Program | DOI | | | |
| Reported Science Research and Investigations (Total) | DOI | 33.293 | 23.737 | 0 |
| Remote Sensing State Grants | DOI | 1.250 | 1.465 | 0 |
| Enhancing Landscape Measurements, Data, and Analysis | DOI | 3.7 | 0 | 0 |
| National Land Use Data Products | DOI | 1.5 | 0 | 0 |
| Natural Capital Accounting | DOI | 3.22 | 0.22 | 0 |
| Baseline Capacity - 2024 Fixed Costs | DOI | 1.036 | 0 | 0 |
| Reported Satellite Operations (Total) | DOI | | | |
| Sustainable Land Imaging Development - Landsat Next | DOI | 103.334 | 91.334 | 0 |
| Baseline Capacity - 2024 Fixed Costs | DOI | 1.047 | 0 | 0 |
| **NCAR** | NSF | 124.59 | Not listed | 116.20 |
| Civil Space Estimate:[*] | | **3,550.074** | **3,178.349** | **3,204.737** |

*AMOUNTS IN MILLIONS.*

---

[*] Calculation includes office totals and individual line items.

## Acronyms

ADS     Bureau of Arms Control, Deterrence, and Stability
AGS     Atmospheric and Geospace Sciences
AST     Office of Commercial Space (FAA)
AST     Astronomical Sciences (NSF)
ATO     Air Traffic Organization
BEA     Bureau of Economic Analysis
CHEXS   Center for High Energy X-ray Science
CISA    Cybersecurity and Infrastructure Security Agency
CJS     Commerce, Justice, and Science (appropriations bill)
CORS    Continuously Operating Reference Stations
CR      Continuing Resolution
CRSRA   Commercial Remote Sensing Regulatory Affairs
DDTC    Directorate of Defense Trade Controls
DHS     Department of Homeland Security
DOC     Department of Commerce
DOD     Department of Defense
DOE     Department of Energy
DOI     Department of the Interior
DOL     Department of Labor
DOS     Department of State
DOT     Department of Transportation
EOP     Executive Office of the President
FAA     The Federal Aviation Administration
FCC     Federal Communications Commission
FY      Fiscal Year
GAGE    Global Aerospace Geodetic Experiment
GNSS    Global Navigation Satellite System
GPS     Global Positioning Satellite
IIJA    Infrastructure Investment and Jobs Act
INSRB   Interagency Nuclear Safety Review Board
ISS     International Space Station
ITA     International Trade Administration
ITAR    International Traffic in Arms Regulations
LEO     Low Earth Orbit
MPS     Mathematical and Physical Sciences
MSaC    NASA's Mission Services and Capabilities
MTS     U.S. Maritime Transportation System
NASA    National Aeronautics and Space Administration

NASS    National Agricultural Statistics Service
NCAR    National Center for Atmospheric Research
NESDIS  National Environmental Satellite, Data, and Information Service
NextGen Next Generation (Communications)
NGF     National Geophysical Facility
NGS     National Geodetic Survey
NIST    National Institute of Standards and Technology
NOAA    National Ocean and Atmospheric Administration
NOIRLab National Optical-Infrared Astronomy Research Laboratory
NOS     National Ocean Service
NRC     Nuclear Regulatory Commission
NSF     National Science Foundation
NTIA    National Telecommunications and Information Administration
OES     Oceans and International Environmental and Scientific Affairs
OIA     Office of International Affairs
OMB     Office of Management and Budget
OSC     Office of Space Commerce
OSHA    Occupational Safety and Health Administration
OSTP    Office of Science and Technology Policy
PHY     Physics
PM      Bureau of Political-Military Affairs
PNT     Positioning, Navigation, and Timing
POE     Port of Entry
SAGE    Seismic Array for Global Earth Observation
SAO     Smithsonian Astrophysical Observatory
SAS/ESC Spectrum Access System and Environmental Sensing Capability
SFS     Space and Flight Support
STEM    Science, Technology, Engineering, and Mathematics
TraCSS  Traffic Coordination System for Space
USDA    United States Department of Agriculture
USGEO   US Group on Earth Observations
USGS    United States Geological Survey
WAAS    Wide Area Augmentation System

## Acknowledgments

The author would like to sincerely thank and acknowledge several individuals for their contributions to this study. Madeline Chang, through the Center for Strategic and International Studies, developed a civil space primer that provided a significant foundation for this paper. Rachel Lindbergh at the Congressional Research Service cited every dollar of NASA funding across several bills and mission name changes over the years. Casey Dreier, from the Planetary Society, served as a peer reviewer to this paper and offered insightful changes that enhanced the value and applicability of this work. Brian Weeden at The Aerospace Corporation contributed to several iterations of this paper with substantial guidance, suggestions, and strategy. The Aerospace Corporation's Technical Publications Department editors suggested enhancements to the paper's applicability and readability.